\documentclass[aps,floatfix,twocolumn,showpacs,amsmath,superscriptaddress]{revtex4}

\usepackage{amssymb}

\usepackage[T1]{fontenc}
\usepackage[latin1]{inputenc}
\usepackage{graphicx}
\usepackage{dcolumn}
\usepackage{bm}

\newcommand{\be}{\begin{eqnarray}}
\newcommand{\ee}{\end{eqnarray}}

\begin{document}


\title{A unique spinodal region in asymmetric nuclear matter}

\author{Jérôme Margueron}
\affiliation{GANIL CEA/DSM - CNRS/IN2P3 BP 5027 F-14076 Caen CEDEX 5, France}
\affiliation{Istituto Nazionale di Fisica Nucleare, Sezione di Pisa, 56100 Pisa, Italy}
\author{Philippe Chomaz}
\affiliation{GANIL CEA/DSM - CNRS/IN2P3 BP 5027 F-14076 Caen CEDEX 5, France}

\date{\today}

\begin{abstract}
Asymmetric nuclear matter at sub-saturation densities is shown to present
only one type of instabilities.
The associated order parameter is dominated by the isoscalar density and so 
the transition is of liquid-gas type. The instability goes in the direction
of a restoration of the isospin symmetry leading to a fractionation 
phenomenon. 
These conclusions are model independent since they can be related to the
general form of the asymmetry energy. They are illustrated using density
functional approaches. 
\end{abstract}



\maketitle

Phase transitions are universal phenomena of matter in interaction. The
coexistence regions which correspond to thermodynamically-forbidden areas 
exhibits general features such as metastabilities or instabilities. 
Since strong interaction between nucleons is of Van der Waals type 
nuclear systems are expected to present a liquid and a gas phase
characterized by their respective densities~\cite{ber80}. Since nucleons can
be either protons or neutrons the transition occurs in a two fluid systems
which 
may lead to a richer phase diagram involving up to two order parameters. It
has been argued that, 
asymmetric nuclear matter (ANM) presents different types of instabilities~: a
broad chemical instability region with the concentration as order parameter
and a narrower domain of mechanical instability for which the total density
plays the role of a second order parameter~\cite{mul95,bao97}. In this
letter, we will show that spinodal instability and phase transition in ANM
involves a unique order parameter which reflects density fluctuation between
the two phases. We will stress that 
the transition induces variations of the concentrations leading to the
isospin fractionation~\cite{bar98} 
similar to the one experimentally observed~\cite{xu00}. These properties and
in fact all the characteristics of the instabilities are independent of
their usual classification as chemical or mechanical instabilities. These
conclusions are related to general properties such as the isospin dependence
of the asymmetry energy. The shape of the spinodal region and the associated
instability times depend upon the model so the observation of spinodal
decomposition may provide constraints on the isospin dependence of the
effective forces.


\smallskip Let us consider ANM characterized by a proton and a neutron
densities $\rho _{i}=$ $\rho _{p}$, $\rho _{n}$. These densities can be
transformed in a set of 2 mutually commuting charges $\rho _{i}=$ $\rho _{1}$%
, $\rho _{3}$ where $\rho _{1}$ is the density of baryons, $\rho _{1}=\rho
_{p}+\rho _{n},$ and $\rho _{3}$ the asymmetry density $\rho _{3}=\rho
_{n}-\rho _{p}$. In infinite matter, the extensivity of the free energy
implies that it can be reduced to a free energy density~: $F(T,V,N_{i})=V%
\mathcal{F}(T,\rho _{i}).$ The system is stable against separation into two
phases if the free energy of a single phase is lower than the free energy in
all two-phases configurations. This stability criterium implies that the free
energy density is a convex function of the densities $\rho _{i}$. A local
necessary condition 
is the positivity of the curvature matrix~: 
\begin{equation}
\left[ \mathcal{F}_{ij}\right] =\left[ \frac{\partial ^{2}\mathcal{F}}{%
\partial \rho _{i}\partial \rho _{j}}|_{T}\right] \equiv \left[ \frac{%
\partial \mu _{i}}{\partial \rho _{j}}|_{T}\right]  \label{eq5}
\end{equation}
where we have introduced the chemical potentials $\mu _{j}\equiv \frac{%
\partial F}{\partial N_{j}}|_{T,V,N_{i}}=\frac{\partial \mathcal{F}}{%
\partial \rho _{j}}|_{T,\rho _{i\not{=}j}}$. In the considered two-fluids
system, the $[\mathcal{F}_{ij}]$ is a $2*2$ symmetric matrix, so it has 2
real eigenvalues $\lambda ^{\pm}$~\cite{bar01}~: 
\begin{equation}
\lambda ^{\pm}=\frac{1}{2}\left( \mathrm{Tr}\left[ \mathcal{F}_{ij}\right]
\pm \sqrt{\mathrm{Tr}\left[ \mathcal{F}_{ij}\right] ^{2}-4\mathrm{Det}\left[ 
\mathcal{F}_{ij}\right] }\right)  \label{eq23}
\end{equation}
associated to 
eigenvectors $\mathbf{\delta \rho }^{\pm}$ defined by ($i\neq j$) 
\begin{equation}
\frac{{\delta \rho }_{j}^{\pm}}{{\delta \rho }_{i}^{\pm}}=\frac{\mathcal{F}%
_{ij}}{\lambda ^{\pm}-\mathcal{F}_{jj}}=\frac{\lambda ^{\pm}-\mathcal{F}_{ii}%
}{\mathcal{F}_{ij}}  \label{eq25}
\end{equation}
Eigenvectors associated with negative eigenvalue indicate the direction of
the instability. It defines a local order parameter since it is the
direction along which the phase separation occurs. 
The eigen values $\lambda $ define sound velocities, $c$, by ${c}^{2}=\frac{1%
}{18m}\rho _{1}\,\lambda .$ In the spinodal area, the eigen value $\lambda $
is negative, so the sound velocity, $c$, is purely imaginary and the
instability time $\tau $ is given by $\tau =d/|c|$ where $d$ is a typical
size of the density fluctuation.

The requirement that the local curvature is positive 
is equivalent to the requirement that both the trace ($\mathrm{Tr}[\mathcal{F%
}_{ij}]=\lambda ^{+}+\lambda ^{-})$ and the determinant ($\mathrm{Det}[%
\mathcal{F}_{ij}]=\lambda ^{+}\lambda ^{-})$ are positive 
\begin{equation}
\mathrm{Tr}[\mathcal{F}_{ij}]\geq 0,\hbox{ and }\mathrm{Det}[\mathcal{F}%
_{ij}]\geq 0  \label{eq6}
\end{equation}
The use of the trace and the determinant which are two basis-independent
characteristics of the curvature matrix clearly stresses the fact that the
stability analysis should be independent of the arbitrary choice of the
thermodynamical quantities used to label the state e.g. $(\rho _{p}$, $\rho
_{n})$ or $(\rho _{1}$, $\rho _{3})$. If Eq.~\ref{eq6} is violated the
system is in the unstable region of a phase transition. Two cases are then
possible~: i) only one eigenvalue is negative and one order parameter is
sufficient to describe the transition or ii) both eigenvalues are negative
and two independent order parameters should be considered meaning that more
than two phases can coexist.

For ANM below saturation density, the case ii) never occurs
since the asymmetry energy has always positive curvature 
($\mathcal{F}_{33}$). Indeed, 
the asymmetry term in the mass formula behave like $(N-Z)^{2}$ times a
positive function of $A$ showing that the dominant $\rho _{3}$ dependence of
the asymmetry potential energy is essentially quadratic and that $\mathcal{F}%
_{33}$ is a positive function of the total density. Recent Bruckner
calculations in ANM~\cite{vid02} have confirmed the positivity of $\mathcal{F%
}_{33}$. They have parameterized the potential energy with the simple form $%
\mathcal{V}(\rho _{1},\rho _{3})=\mathcal{V}_{0}(\rho _{1})\rho _{1}^{2}+%
\mathcal{V}_{1}(\rho _{1})\rho _{3}^{2}$ with $\mathcal{V}_{0}/\mathcal{V}%
_{1}\sim -3$ and $\mathcal{V}_{0}<0$. This is also true for effective forces
such as Skyrme forces. For example, the simplest interaction with a constant
attraction, $t_{0}$, and a repulsive part, $t_{3}\rho _{1}$, leads to $%
\mathcal{V}(\rho _{1},\rho _{3})=(3\rho _{1}^{2}-\rho _{3}^{2})\beta (\rho
_{1})$ with $\beta (\rho _{1})=(t_{0}+t_{3}/6\rho _{1})/8$. The function $%
\beta (\rho _{1})$ is negative below saturation density, hence the
contribution of the interaction to $\mathcal{F}_{33}$ in the low density
region is always positive.

These arguments show that, below saturation density, the $\rho _{3}$
curvature, $\mathcal{F}_{33}$, is expected to be positive for all
asymmetries. Since the curvature in any direction, $\mathcal{F}_{ii}$,
should be between the two eigenvalues $\lambda ^{-}\leq \mathcal{F}_{ii}\leq
\lambda ^{+}$ we immediately see that if $\mathcal{F}_{33}$ is positive one
eigen curvature at least should remain positive.
In fact for all models we have studied $\mathcal{F}_{33}$ appears to be
always large enough so that the trace is always positive demonstrating that $%
\lambda ^{+}>0$. Since $\mathrm{Tr}[\mathcal{F}_{ij}]=\mathcal{F}_{nn}+%
\mathcal{F}_{pp}$, this can be related to the positivity of the Landau
parameter $\mathcal{F}_{nn}$ and $\mathcal{F}_{pp}$.

\smallskip The large positive value of $\mathcal{F}_{33}$ also indicates
that the instability should remain far from the $\rho _{3}$ direction i.e.
it should involve total density variation and indeed we will see that in all
models and for all asymmetry the instability direction hardly deviates from
a constant asymmetry direction $(\delta \rho _{3}\ll \delta \rho _{1})$.
This isoscalar nature of the instability can be understood by looking at the
expression of the eigen-modes in the $(\rho _{n};\rho _{p})$ coordinates.
Since $\lambda ^{-}\leq \mathcal{F}_{ii}\leq \lambda ^{+}$, 
the differences $\lambda ^{-}-\mathcal{F}_{ii}$ 
is always negative demonstrating, using Eq.~\ref{eq25}, that the
instability is of isoscalar type, if the Landau parameter $\mathcal{F}_{np}$
is negative. 
Hence, there is a close link between the isoscalar nature of the instability
and the attraction of the proton-neutron interaction~\cite{bar01}.


\begin{figure}[htbp]
\center
\includegraphics[scale=0.7]{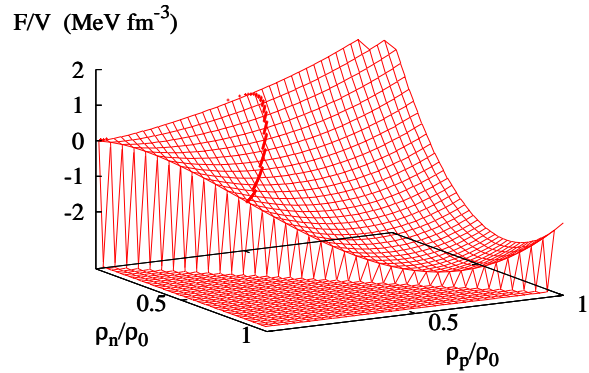}
\caption{This figure represents the energy surface as a function of the
densities $\rho_n$ and $\rho_p$ for the SLy230a interaction. The contour
delimitate the spinodal area. }
\label{fig1}
\end{figure}

\smallskip 
To illustrate the above results we will now use density
functional formalism using Skyrme~\cite{sky56} and Gogny~\cite{gog75}
effective forces. 
It should be noticed, that the extraction of the sound velocity corresponds
to a random phase approximation (RPA) which goes beyond the mean field
approximation. 
We will focus on the isospin degree of freedom for which RPA approaches can
be considered as good approximations and discuss both zero and finite
temperature in ANM.

We represent in Fig.~\ref{fig1} the energy surface as a function of $\rho
_{n}$ and $\rho _{p}$, deduced from SLy230a Skyrme interaction~\cite{cha97}.
In the symmetric case ($\rho _{n}=\rho _{p}$), one can see the
negative curvature of the energy which defines the spinodal area, whereas in
pure neutron matter ($\rho _{p}=0$), no negative curvature and so no
spinodal instability are predicted. We can also notice that the isovector
density dependence is almost parabolic illustrating the positivity 
of $\mathcal{F}_{33}$.

\begin{figure}[tbph]
\center
\includegraphics[scale=0.3]{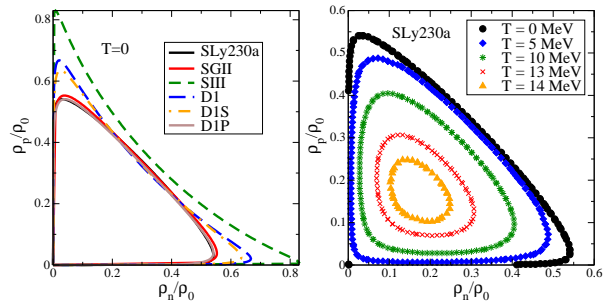}
\caption{This two figures are a projection of the spinodal contour in the
density plane : left, for Skyrme (SLy230a~\protect\cite{cha97}, SGII~ 
\protect\cite{ngu81}, SIII~\protect\cite{bei75}) and Gogny models (D1~%
\protect\cite{gog75}, D1S~ \protect\cite{ber91}, D1P~\protect\cite{far99}) ;
right, temperature dependence of the spinodal zone computed for the SLy230a
case.}
\label{fig2}
\end{figure}

The spinodal contours predicted by several models exhibit important
differences (see Fig.~\ref{fig2}). In the case of SLy230a force (as well
as SGII, D1P), the total density at which spinodal instability appears
decreases when the asymmetry increases whereas for SIII
(as well as D1, D1S) it increases up to large asymmetry and finally
decreases. 
Considering the implicit equation of the spinodal limit $g(\rho _{1},\rho
_{3})\equiv \mathrm{Det}[\mathcal{F}_{ij}]=0$, we can show that the
curvature of the spinodal around the symmetry is related to $\partial ^{2}g/%
\partial \rho _{3}^{2}(\rho _{3}=0)$. The isospin symmetry 
impose that $\partial g/\partial \rho _{3}(\rho _{3}=0)=0$ and one can show
that 
\begin{eqnarray}
\frac{\partial ^{2}g}{\partial \rho _{3}^{2}}|_{\rho _{3}=0} &=&\mathrm{Det}%
\left[ \frac{\partial \mathcal{F}_{ij}}{\partial \rho _{3}}|_{\rho
_{3}=0}\right]   \nonumber \\
&+&(\mathcal{F}_{1133}\mathcal{F}_{33}+\mathcal{F}_{3333}\mathcal{F}_{11}+%
\mathcal{F}_{1333}\mathcal{F}_{13}).
\end{eqnarray}
Assuming that the forth derivatives of the free energy are negligible
compare to the third one, the curvature of $g$ is mainly given by $\mathrm{%
Det}[\partial \mathcal{F}_{ij}/\partial \rho _{3}]$ where $\partial \mathcal{%
F}_{ij}/\partial \rho _{3}$ is nothing but the curvature matrix of $\partial %
F/\partial \rho _{3}$, i.e. the asymmetry density dependence of the free
energy. 
We observe that all forces which fulfill the global requirement that they
reproduce symmetric nuclear matter (SNM) equation of state as well as the
pure neutron matter calculations, leads to the same curvature of the
spinodal region.

The temperature dependence of the spinodal contour can be appreciated in the
right part of Fig.~\ref{fig2}). As the temperature increases the spinodal
region shrinks up to the critical temperature for which it is reduced to SNM
critical point. However, up to a rather high temperature ( $5$ to $10$ $%
\mathrm{MeV}$) the spinodal zone remains almost identical to the zero
temperature one.

\begin{figure}[htbp]
\center
\includegraphics[scale=0.3]{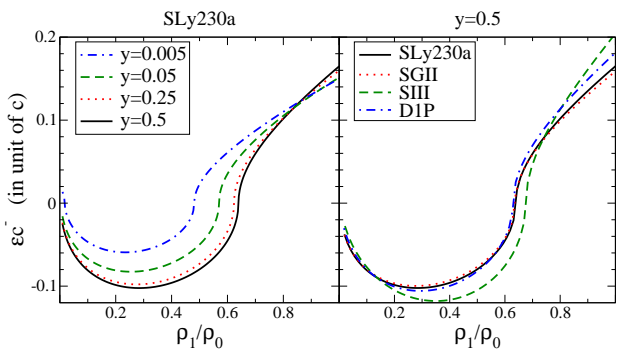}
\caption{Sound velocity as a function of the density. Negative values of
sound velocity means that it is imaginary ($\epsilon=i$ is $c^-$ imaginary).
On the left, we have changed the asymmetry parameter and fixed the model
(SLy230a), on the right, we have changed the models and fixed
$y=\rho_p/\rho_1$=0.5. }
\label{fig5}
\end{figure}

We show in the right part of Fig.~\ref{fig5} the sound velocity in SNM as
a function of the density 
for several forces. When we enter into
the spinodal area, the sound velocity becomes purely imaginary. 
The different forces predict different instability time. However, for the
set of forces fitted to reproduce the symmetric and the pure neutron matter
calculation, we observe a convergence of prediction as we already observed
on the spinodal boundaries. 
In the left part of Fig.~\ref{fig5} we can appreciate the reduction of
the instability when we go away from SNM. However, large asymmetries are
needed to induce a sizable effect.

\begin{figure}[htbp]
\center
\includegraphics[scale=0.25]{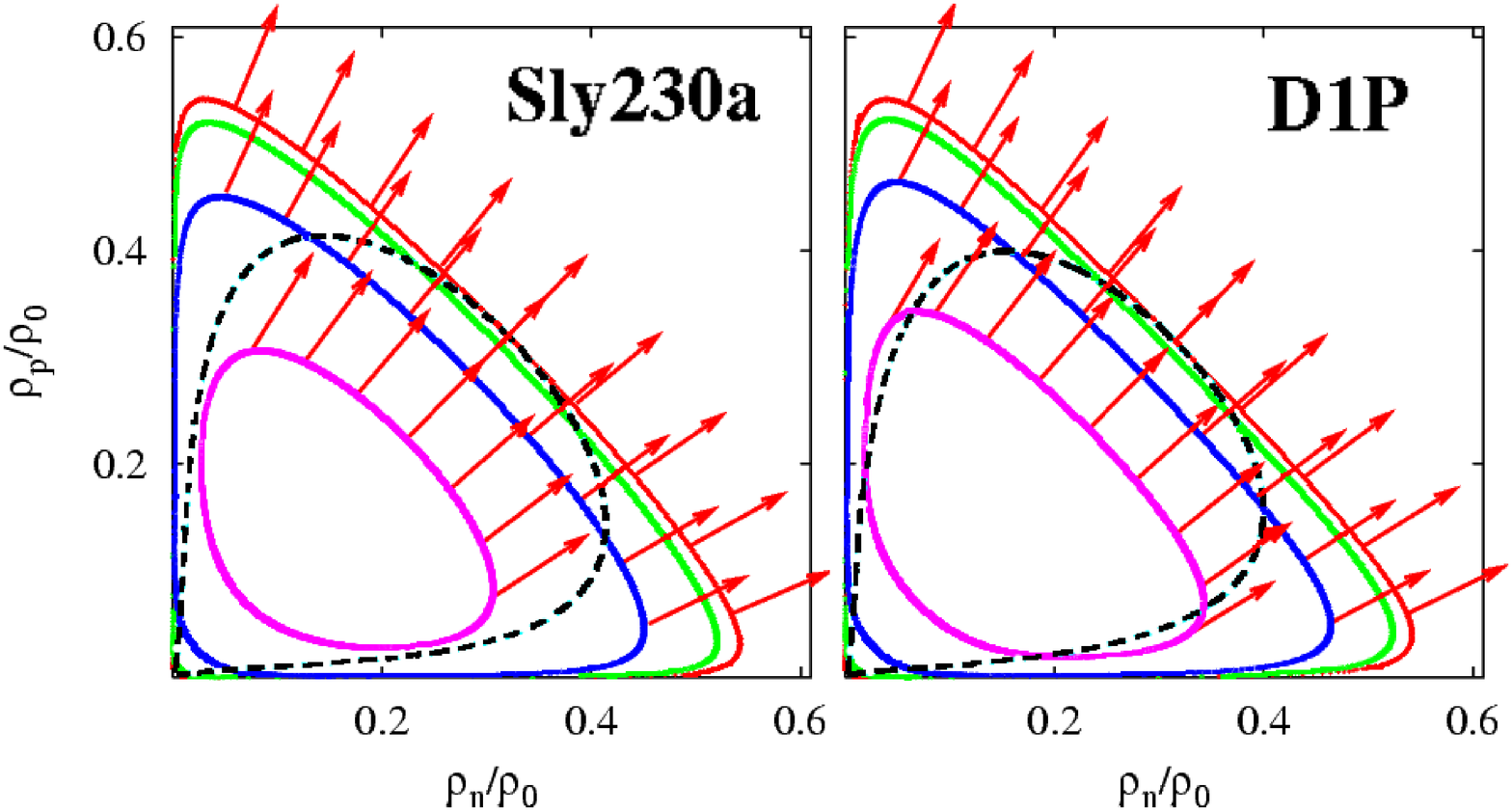}
\caption{This is the projection of the iso-eigen values on the density plane
for Slya (left) and D1P (right). The arrows indicate the direction of
instability. The mechanical instability is also indicated (dotted line).}
\label{fig3}
\end{figure}

Various contour of equal imaginary sound velocity are represented in 
Fig.~\ref{fig3} for SLy230b and D1P interactions. The more internal curve
correspond to the sound velocity $i0.09c$, after comes $i0.06c$, $i0.03c$
and finally 0, the spinodal border. 
For these two recent forces which takes into account pure neutron matter
constraints, the predicted instability domains are rather similar. We
observe that in almost all the spinodal region the sound velocity is larger
than $0.06c$. 

\begin{figure}[tbph]
\center
\includegraphics[scale=0.2]{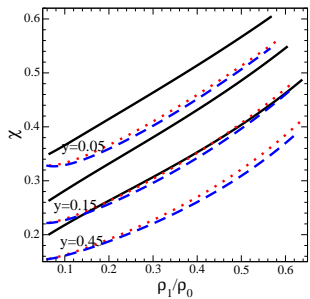}
\caption{Angle of $\mathbf{\delta \rho }^{-}$ with the isoscalar axis
normalized by the angle between the $y$=const line and the isoscalar axis 
(denoted $\chi$)
for various values of $y$ noted on the figure and for 3 models (solid:
SLy230a, dotted: D1P, dashed: SGII).}
\label{fig6}
\end{figure}

\smallskip Let us now focus on the direction of the instability. 
If $\mathbf{\delta \rho }^{-}$ is along $y=\rho_p/\rho_1$=const
then the instability does not change the proton fraction. 
For symmetry reasons pure isoscalar $(\delta \rho _{3}=0)$ and isovector $%
(\delta \rho _{1}=0)$ modes appears only for SNM so it is interesting to
introduce 
a generalization of isoscalar-like and isovector-like modes by considering
if the protons and neutrons move in phase ($\delta \rho _{n}^{-}\delta \rho
_{p}^{-}>0$) or out of phase ($\delta \rho _{n}^{-}\delta \rho _{p}^{-}<0$).
Fig.~\ref{fig3} shows the direction of instabilities along the spinodal
border and some iso-instability lines. We observed that instability is
always almost along the $\rho _{1}$ axis meaning that it is dominated by
total density fluctuations even for large asymmetries. Fig.~\ref{fig6}
presents the angle of the eigen state $\mathbf{\delta \rho }^{-}$ with the
isoscalar axis normalized by the angle between the $y$=const line and the
isoscalar axis (denoted $\chi$) for 3 models (D1P, SGII, SLy230a). 
We can see that this quantity is in between 0 and 1, so that the instability
direction is between the $y$=const line and the $\rho _{1}$ direction. 
This shows that the unstable direction is of isoscalar nature as expected
from the attractive interaction between proton-neutron. 
The total density is therefore the dominant contribution to the order
parameter showing that the transition is between two phases having different
densities (i.e. liquid-gas phase transition). 
The angle with the $\rho _{1}$ axis is almost constant along a constant $y$
line. This means that as the matter enters in the spinodal zone and then
dives into it, there are no dramatic change in the instability direction
which remains essentially a density fluctuation. Moreover, the unstable
eigenvector drives the dense phase (i.e. the liquid) towards a more
symmetric point in the density plane. By particle conservation, the gas
phase will be more asymmetric leading to the fractionation phenomenon. 
Those results are in agreement with recent calculation for ANM~\cite{bar01} 
and nuclei~\cite{col02}.

We want to stress that those qualitative conclusions are very robust and
have been reached for all the Skyrme and Gogny forces we have tested (SGII,
SkM$^{*}$, RATP, D1, D1S, D1P,...) including the most recent one (SLy230a,
D1P) as well as the original one (like SIII, D1). 

\smallskip A different discussion can be found in the literature~\cite
{mul95,bao97}.
Therein, a chemical and mechanical stability conditions 
\begin{equation}
\frac{\partial \mu _{p}}{\partial y}|_{T,P}>0\hspace{0.5cm}\hbox{ and }%
\hspace{0.5cm}\frac{\partial P}{\partial \rho _{1}}|_{T,y}>0
\label{EQ:Mechanical}
\end{equation}
are introduced and two types of spinodal regions are defined. This leads to
the idea that two order parameters should be introduced~: the concentration
in the chemical instability zone and the baryon density in the mechanical
instability region. To try to connect the eigenstate analysis with this 
discussion one can use the relations 
\begin{equation}
\rho _{n}\,\mathrm{Det}[\mathcal{F}_{ij}]=\frac{\partial \mu _{p}}{\partial y%
}|_{T,P}\,\,\frac{\partial P}{\partial \rho _{1}}|_{T,y}  \label{eq9}
\end{equation}
Comparing expression (\ref{eq9}), with the identity $\mathrm{Det}[\mathcal{F}%
_{ij}]=\lambda ^{+}\lambda ^{-}$ one can be tempted 
to relate separately $\frac{\partial \mu _{p}}{\partial y}|_{T,P}$ and $%
\frac{\partial P}{\partial \rho _{1}}|_{T,y}$ to the two eigenvalues $%
\lambda ^{+}$ and $\lambda ^{-}$. This is indeed correct in SNM~\cite{bar01}
but theses relations break down in ANM. For instance, 
$\frac{\partial P}{\partial \rho _{1}}|_{T,y}=\rho _{1}\frac{\partial ^{2}%
\mathcal{F}}{\partial \rho _{1}^{2}}|_{T,y}
$ 
is nothing but the curvature of the free energy in the particular direction
of constant proton fraction and this direction has 
no reason to be an eigenvector direction, except in SNM. 
Considering Eq.~\ref{eq9} only the product $\frac{\partial \mu _{p}}{%
\partial y}|_{T,P}\,*\,\frac{\partial P}{\partial \rho _{1}}|_{T,y}$ should
be used to spot the instability region. 
Eq.~\ref{eq9} shows that at the onset of instability where $\mathrm{%
Det}[\mathcal{F}_{ij}]=0$ the product should vanish. On the spinodal
border, $\frac{\partial P}{\partial \rho _{1}}|_{T,y}$ vanishes only if the
direction of instability is the $y$=const line. This happens only for SNM
therefore in general $\frac{\partial \mu _{p}}{\partial y}|_{T,P}$ should
vanish first irrespectively of the actual nature of the instability. In
particular, the previous eigen vector analysis shows that in the so called
chemical region not only the concentration fluctuations, $\delta y$, are
amplified but mainly the total density one, $\delta \rho _{1}$. When
entering in the mechanical spinodal zone (shown in Fig.~\ref{fig3}) nothing
special happens : the instability strength evolves smoothly (see also
Fig.~\ref{fig5}) and the associated vector keeps pointing in the same 
isoscalar-like direction (see Fig.~\ref{fig6}). Consequently, the chemical
and mechanical stability conditions should not be considered separately but
combined into 
the determinant of the curvature
matrix.

In this paper, we have shown that ANM does not present two types of spinodal
instabilities, a mechanical and chemical, but only one which is dominantly
of isoscalar nature as a consequence of the negativity of the Landau
parameter $\mathcal{F}_{np}$. This general property can be linked to the
positivity of the symmetry energy curvature $\mathcal{F}_{33}$. This means
that the instability is always dominated by density fluctuations and so can
be interpreted as a liquid-gas separation. The instabilities tend to restore
the isospin symmetry for the dense phase (liquid) leading to the
fractionation of ANM.
We have shown that changing the asymmetry up to $\rho _{p}<3\rho _{n}$ does
not change quantitatively the density at which instability appears, neither
the imaginary sound velocity compared to those obtained in SNM. All the
above results 
are not qualitatively modified by the temperature which mainly introduce a
reduction of the spinodal region up to the SNM critical point where it
vanishes. The quantitative predictions concerning the shape of the spinodal
zone as well as the instability times depends upon the chosen interaction
but converge for the various forces already constrained to reproduce the
pure neutron matter calculation.

We want to thank Bao-An Li and V.Baran for interesting discussion during the
preparation of this manuscript.

\end{document}